# High-resolution x-ray telescopes

Stephen L. O'Dell[*a], Roger J. Brissenden[b], William N. Davis[b], Ronald F. Elsner[a], Martin Elvis[b], Mark Freeman[b], Terry Gaetz[b], Paul Gorenstein[b], Mikhail V. Gubarev[a], Diab Jerius[b], Michael Juda[b], Jeffery J. Kolodziejczak[a], Stephen S. Murray[b,c], Robert Petre[d], William Podgorski[b], Brian D. Ramsey[a], Paul B. Reid[b], Timo Saha[d], Daniel A. Schwartz[b], Susan Trolier-McKinstry[e], Martin C. Weisskopf[a], Rudeger H. T. Wilke[e], Scott Wolk[b], and William W. Zhang[d]

[a] NASA Marshall Space Flight Center, Space Science Office, Huntsville, AL 35812, USA
[b] Smithsonian Astrophysical Observatory, 60 Garden St., Cambridge, MA 02138, USA
[c] Johns Hopkins University, Department of Physics & Astronomy, Baltimore, MD 21218, USA
[d] NASA Goddard Space Flight Center, Greenbelt, MD 20771, USA
[e] Pennsylvania State University, Materials Research Institute, University Park, PA 16802, USA

**ABSTRACT**

High-energy astrophysics is a relatively young scientific field, made possible by space-borne telescopes. During the half-century history of x-ray astronomy, the sensitivity of focusing x-ray telescopes—through finer angular resolution and increased effective area—has improved by a factor of a 100 million. This technological advance has enabled numerous exciting discoveries and increasingly detailed study of the high-energy universe—including accreting (stellar-mass and super-massive) black holes, accreting and isolated neutron stars, pulsar-wind nebulae, shocked plasma in supernova remnants, and hot thermal plasma in clusters of galaxies. As the largest structures in the universe, galaxy clusters constitute a unique laboratory for measuring the gravitational effects of dark matter and of dark energy. Here, we review the history of high-resolution x-ray telescopes and highlight some of the scientific results enabled by these telescopes. Next, we describe the planned next-generation x-ray-astronomy facility—the *International X-ray Observatory* (IXO). We conclude with an overview of a concept for the next next-generation facility—Generation X. The scientific objectives of such a mission will require very large areas (about 10000 m$^2$) of highly-nested lightweight grazing-incidence mirrors with exceptional (about 0.1-arcsecond) angular resolution. Achieving this angular resolution with lightweight mirrors will likely require on-orbit adjustment of alignment and figure.

## 1. INTRODUCTION

Owing to the opacity of the earth's atmosphere, x-ray astronomy is necessarily a "space-age" science. X-ray detectors aboard sub-orbital rockets detected x-ray emission from the sun in 1949[1] and discovered the first (extra-solar) cosmic x-ray source in 1962[2]. In the intervening years, technological improvements—most importantly, the development of high-resolution (focusing) x-ray telescopes—has improved the sensitivity for detecting cosmic x-ray sources by a factor[3] of ten billion ($10^{10}$). The crowning achievement in x-ray telescopes is the *Chandra X-ray Observatory*[4], which provides sub-arcsecond imaging and high-resolution dispersive spectroscopy.

Two factors—angular resolution and collecting area—are paramount in the imaging performance of a telescope. While some current and planned x-ray telescopes offer more collecting area than *Chandra*, none rivals its angular resolution. Indeed, owing to mass and envelope constraints attendant to space-borne telescopes, improving angular resolution and increasing collecting area are often conflicting goals: Reducing mass generally decreases stiffness, which increases susceptibility to distortion and thus to image degradation. Both traditional and state-of-the-art approaches have difficulty in overcoming this obstacle to realizing very-large-area, high-resolution x-ray telescopes.

---

[*] Contact author: Steve.O'Dell@NASA.gov; phone +1 256-961-7776; fax +1 256-961-7522; wwwastro.msfc.nasa.gov.

A novel approach to addressing this problem is to develop adaptive x-ray optics for space-borne telescopes. The objective is to make in-space corrections of figure and alignment errors to optimize the angular resolution of the x-ray telescope. The expectation is that initial and occasional—not real-time—adjustments would be required. Due to the geometry of highly-nested grazing-incidence optics, actuation of the mirrors requires an approach differing from that used for normal-incidence telescopes. Thus far, two groups are conducting research into adjustable x-ray telescopes: The (UK) Smart X-ray Optics consortium[5,6,7] and the (US) Generation-X Concept team[8,9,10,11,12]. These Proceedings report current research into adjustable x-ray telescopes and related technologies, which is being conducted by the Smart X-ray Optics consortium[13,14], by the Generation-X team[15,16,17,18,19], and by the INAF-Brera x-ray optics group[20,21].

The intent of the present paper is to "set the stage" for these more detailed technical reports, by describing the context and the motivation for this research. Section 2 provides background information on x-ray astronomy and telescopes; Section 3 offers a brief tour of the x-ray universe at high angular resolution; and Section 4 describes past, current, and future high-resolution x-ray observatories. To conclude, Section 5 sketches some of the challenges in developing a very-large-area, high-resolution x-ray telescope for a next next-generation x-ray observatory.

## 2. BACKGROUND INFORMATION

This section provides background information on x-ray astronomy and x-ray telescopes. Section 2.1 summarizes of the early history of x-ray astronomy. Next, Section 2.2 briefly describes grazing-incidence reflection and how it affects telescope design, while Section 2.3 discusses the Wolter-1-like geometry of grazing-incidence x-ray telescopes. Section 2.4 defines the key metrics that characterize the imaging performance of a focusing x-ray telescope.

### 2.1. Early history of x-ray astronomy

In 1949, Friedman et al.[1] used photon counters aboard a V-2 rocket to detect x radiation from the solar corona. This confirmed less-direct measurements by colleagues at the US Naval Research Laboratory, in V-2 rocket flights during 1948–1949. In 1962, using 3 large-area Geiger counters aboard an Aerobee rocket, Giacconi et al.[2] discovered the first (extra-solar) cosmic x-ray source, thus initiating the field of x-ray astronomy. The early x-ray-astronomy satellites to follow generally employed proportional counters with mechanical collimators to locate cosmic x-ray sources. These missions included[22,23,24,25] NASA's *Uhuru* (1970–1973), the Astronomy Netherlands Satellite (ANS, 1974–1976), the UK's Ariel V (1974–1980), NASA's Small Astronomy Satellite 3 (SAS-3, 1975–1980), and NASA's High-Energy Astronomy Observatory 1 (HEAO-1, 1977–1979).

In 1960, Giacconi and Rossi[26] had designed a single-reflection grazing-incidence concentrator and mentioned use of two-reflection grazing-incidence optics for x-ray telescopes, using a ("Wolter-I", §2.3) configuration similar to that suggested by Wolter[27] for x-ray microscopes. This led[3] to a 1963 unsolicited proposal[28] to NASA for an orbiting x-ray observatory utilizing (Wolter-I) grazing-incidence mirrors. By 1965, Giacconi et al.[29] had obtained x-ray photographic images of the sun, using 3 small (two-reflection) grazing-incidence telescopes[30] aboard an Aerobee rocket. Over the following 10 years, several sub-orbital rockets[31,32], the Orbiting Solar Observatory 4[33] (OSO-4, 1967-1969), and the space station Skylab[34,35] (1973-1974) carried small (Wolter-I-like) x-ray telescopes for solar observations.

The launch of NASA's High-Energy Astronomy Observatory 2 (HEAO-2, 1978-1981)—the *Einstein Observatory*[36]—heralded a new era[37,38] in x-ray astronomy. *Einstein* was the first astronomical observatory employing a focusing x-ray telescope. With a collecting area two-orders of magnitude larger than solar x-ray telescopes, *Einstein* enabled high-resolution imaging of cosmic x-ray sources from an orbiting observatory, with a sensitivity (§2.4) that allowed detection of x radiation from many classes of astronomical objects. This elicited the interest of the broader astronomical community in x-ray data. The HEAO-2 leadership built upon this interest to initiate a guest-observer program, which served as a model for general-user astronomical facilities operated by NASA, by ESA, and by JAXA.

The *Einstein Observatory* was a scaled-down version of the 1.2-m-diameter x-ray telescope originally proposed[28] by Giacconi and Gursky in 1963. In 1999, NASA launched a full-scale version—the *Chandra X-ray Observatory*[4]. With its unprecedented sub-arcsecond x-ray imaging[39], *Chandra* provides exquisite x-ray images (§3) comparable in resolution to those obtained in the visible and infrared bands. Owing largely to its exceptional angular resolution, *Chandra* is 3 orders of magnitude more sensitive than *Einstein*, 7 orders better than the first (1965) focusing x-ray telescope[29], and 10 orders better than the (1962) instrument[2] that detected the first extra-solar x-ray source.

## 2.2. Grazing-incidence reflection

Reflection of x rays relies upon the principle of near-total external reflection[a]. Thus, for a given x-ray photon energy $E$, there is a critical grazing angle $\vartheta_c(E)$ below which the reflectance is high:

$$\sin\vartheta_c(E) = \frac{E_p}{E}\sqrt{\frac{f_1(E)}{Z}} = \frac{(0.029\,\text{keV})}{E}\sqrt{\frac{f_1(E)}{A}\rho} \approx \frac{(0.082\,\text{keV})}{E}\sqrt{\frac{\rho}{20}}\sqrt{\frac{f_1(E)}{Z}}, \qquad (1)$$

where $E_p = \hbar\omega_p$ is the (free-electron) plasmon energy, proportional to the plasma frequency $\omega_p$ of the medium. In determining the x-ray reflectance, the relevant parameters of the medium are its specific density $\rho$, atomic mass $A$, atomic number $Z$, and atomic scattering factor $f_1(E) + i f_2(E)$. The above approximation sets $Z \approx 0.40 A$, which is typical of high-$Z$ optical coatings (e.g., gold, platinum, iridium, and tungsten) having specific densities of about 20.

The critical grazing angle is then roughly inversely proportional to x-ray energy[b], with a value of about 8 mrad ≈ 30′ near 10 keV, for typical (high-Z, high-density) optical coatings. Consequently, science requirements specifying the x-ray energy range significantly influence the range of cone angles $\alpha$ used in designing a grazing-incidence telescope.

Thus far, all high-resolution x-ray telescopes (§4) have covered only the soft-x-ray band—i.e., up to about 10 keV. At lower angular resolution, balloon-borne hard-x-ray telescopes[40,41] have demonstrated the capability for focused imaging of cosmic sources up to several tens of keV. In addition, two satellites currently in preparation will carry hard-x-ray telescopes[42,43]. Although their angular resolutions are moderate (of order arcminute), they are much more sensitive (§2.4) than previous (non-focusing) hard-x-ray instruments. In practice, there is an upper limit to the energy range of a grazing-incidence telescope, in that the ratio of aperture to mirror surface area is $\sin\alpha \propto 1/E_{\max}$. The use of graded multilayers for hard-x-ray telescopes[44,45,46,47] somewhat relaxes, but does not eliminate, this limitation.

## 2.3. Wolter-I-like geometry

For grazing incidence, true imaging requires an even number of reflections. A grazing-incidence paraboloid focuses an on-axis point source; however, any single-reflection grazing-incidence optic produces severe aberrations for off-axis or extended sources: Thus, it is effectively a concentrator rather than an imaging telescope. Figure 1 illustrates the operation of a focusing grazing-incidence telescope that provides true imaging through two reflections.

Thus far, all high-resolution x-ray telescopes utilize a geometry similar to the displayed Wolter type-I geometry[27], with two grazing-incidence reflections from the *inner* surfaces of near-cylindrical mirrors. For most astronomical x-ray telescopes—including the *Chandra* High-Resolution Mirror Assembly (HRMA)—primary and secondary mirrors are paraboloid and hyperboloid, respectively. This prescription and the similar Wolter–Schwarzschild prescription[48] result in aberration-free on-axis imaging, which optimizes narrow-field imaging. Other prescriptions (hyperboloid-hyperboloid, polynomial, etc.) optimize wide-field imaging with a Wolter-I-like geometry, but at the expense of on-axis performance. A few extreme-ultraviolet (EUV) telescopes utilize the Wolter type-II geometry, for which the second grazing-incidence reflection is from the *outer* surface of a hyperboloid, to feed high-resolution spectrometers. However, for large-area telescopes, the Wolter-I-like geometry has decisive advantages:

1. It is more compact for a given grazing angle, as each reflection converges and deflects x rays in the same direction.
2. It accommodates a high degree of nesting, more effectively filling the available aperture area.
3. It is less difficult to fabricate, align, and assemble.

---

[a] Within geometric optics, grazing-incidence reflection provides achromatic focusing. Focusing by diffractive optics—e.g., Laue/Bragg lens or Fresnel zone plates—is highly chromatic and thus less suitable for concentrating or imaging broad-band sources.

[b] The real part of the atomic scattering factor $f_1(E) \to Z$ for large $E$. Thus, in the hard-x-ray band (10–100 keV, say), $\sqrt{f_1/Z} \approx 1$, away from atomic edges (where dispersion is anomalous). Over the soft x-ray band (0.1–10 keV, say), $\sqrt{f_1/Z}$ increases from about 0.5 (for the high-$Z$ optical coatings) to 1, again away from atomic edges.

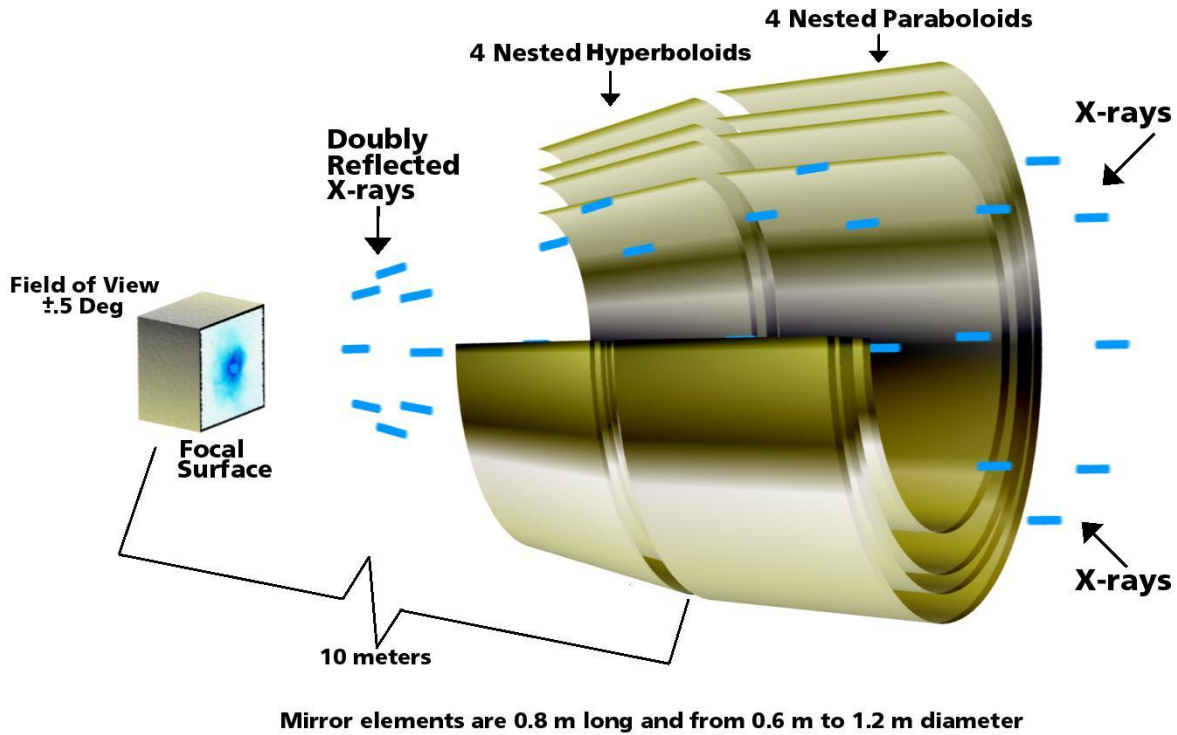

Figure 1: Cut-away schematic of the *Chandra X-ray Observatory*'s x-ray telescope. Four nested, co-axial, confocal, grazing-incidence mirror pairs focus cosmic x rays onto the telescope's focal surface. All past, present, and planned high-resolution x-ray telescopes utilize this basic (Wolter-I-like) geometry—albeit with differing mirror prescriptions, dimensions, and number of shells. [Credit: NASA/CXC/ D. Berry]

**2.4. Performance metrics**

Two key metrics characterize the optical performance of an x-ray telescope. These govern its imaging capability and its sensitivity for detecting sources and accurately measuring their spectral flux. Section 2.4.1 discusses the angular resolution; Section 2.4.2, the collecting area. Section 2.4.3 elaborates on the factors that determine detection sensitivity.

**2.4.1. Angular resolution**

The point spread function (PSF) fully describes the image of a point source as a function of the photon's energy and direction with respect to the optical axis. For a high-resolution x-ray telescope[c], the PSF comprises a compact core (determined by alignment errors, surface figure deviations from the optical prescription, and aberrations inherent in the prescription) and an extended halo or wings (caused by diffractive scattering by surface micro-roughness and by particulates on the surface). The PSF core is then essentially geometric in origin and, thus, approximately independent of photon energy (achromatic).[d] On the other hand, the PSF wings are diffractive in origin and, thus, dependent upon photon energy (chromatic). Hence, minimizing the PSF wings typically is an issue of limiting mirror micro-roughness (e.g., by super-polishing to about 0.3-nm micro-roughness or better) and of controlling particulate contamination[49,50]. Thus, the processes for minimizing the PSF wings are largely decoupled from those for optimizing the PSF core.

---

[c] The relevant unit for the angular resolution of precision x-ray telescopes is an arcsecond (″): $1″ = 4.848$ μradian.

[d] Large-area x-ray telescopes nest multiple co-aligned, co-axial, grazing-incidence (primary and secondary) mirror pairs in order to utilize more efficiently the available aperture (§2.3). Owing to the sensitivity of reflectance to grazing angle (§2.2, §2.4.2), the energy-dependent response of each mirror shell depends upon its mean axial slope (cone angle) $\alpha$. Consequently, although the PSF core of each mirror shell is nearly energy independent, the PSF core of an ensemble of mirror shells can be energy dependent—even in the geometric-optics limit.

For (focusing) x-ray telescopes, the standard metric for specifying angular resolution is the half-power diameter (HPD), which is also called the "half-energy width" (HEW). This is the angular diameter of the image of a point source, which contains half the flux (at a given energy) focused by the telescope. From the standpoint of detecting and measuring sources with an x-ray telescope, the HPD proves more useful than other imaging metrics—e.g., full width at half maximum[e] (FWHM) and root-mean-of squares (RMS) image blur[f]. However, the RMS is useful in formulating imaging error budgets for the geometric-optics terms that govern the PSF core[g].

Figure 2 illustrates an obvious advantage of finer angular resolution in imaging an extended x-ray source and elucidating its structure. The four panels show x-ray images of the Crab Nebula (§3) obtained with the four highest resolution x-ray observatories (§4): XMM-*Newton*, the *Einstein Observatory*, the *Röntgen Satellit* (ROSAT), and the *Chandra X-ray Observatory*.

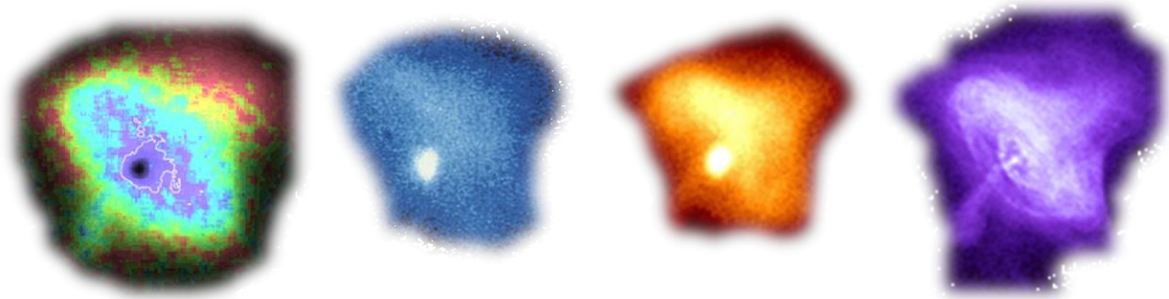

Figure 2: Comparison of x-ray images of the Crab Nebula obtained with high-resolution x-ray telescopes. From left to right, half-power-diameter (HPD) resolutions are approximately 15″, 10″, 5″, and 0.6″. The highest resolution image clearly shows much more structure, including an "inner ring" with semi-major and semi-minor axes of 7″ and 14″.

The other advantage of finer angular resolution is in detecting and measuring the flux from faint unresolved sources. If the angular resolution is good, there is a low probability for a background event—i.e., one not associated with the x-ray source—to occur within the image of an unresolved ("point") source. This drives the background level in a detection cell to very low levels, proportional to the HPD squared. This is the primary reason why focusing x-ray telescopes are so much more sensitive than non-focusing telescopes, which use, e.g., mechanical collimators or coded apertures to determine source positions. During quiescent periods, the Chandra CCDs have a background rate, within the 0.3-arsec$^2$ area (approximately 1 CCD pixel) defined by the telescope's HPD, of only 1 event/keV per year!

---

[e] Although full width at half maximum (FWHM) is a standard measure of angular resolution for normal-incidence optics, it is generally less informative than HPD for grazing-incidence optics. In the grazing-incidence geometry, axial-slope deviations alone form a $(1/\vartheta)$ cusp in the PSF, which out-of-roundness errors (in cone angle and radius) and misalignments broaden. Consequently, an azimuthally symmetric mirror shell has a small ("very good") FWHM; however, that FWHM would contain very little power if the axial-slope deviations were large.

[f] Using the root-mean-of squares (RMS) blur (radius or diameter) can disproportionately weight a small fraction of the power that falls very far from the image core—due to diffractive-scattering wings or to a very large figure error over a small fraction of the mirror. From the standpoint of detecting or measuring a cosmic x-ray source, one cares more about the radiation in or near the image core than about the radiation far outside the core. However, minimizing the PSF wings is still important: X rays scattered from bright sources limit the sensitivity for detecting faint sources in their vicinity.

[g] Because statistical variances of independent processes rigorously add, combining RMS error terms into a total RMS image blur simply requires a root sum of squares (RSS) of the individual error terms. Note that, *for Gaussian distributed errors*, HPD = 0.675 RMSD = 1.35 RMSR and FWHM = 1.177 RMSD = 2.355 RMSR, where RMSR = $\sigma$ is the standard deviation of the distribution.

### 2.4.2. Collecting area

The standard metric for specifying the collecting area of an x-ray telescope is the on-axis, energy-dependent, effective area $A_{\text{eff}}(E)$. For a nested system of mirror shells, the total effective area is just the sum of the effective areas of the individual mirror shells $m$:

$$A_{\text{eff}}(E) = \sum_m A_{\text{eff},m}(E) \cong \sum_m A_{\text{ap},m}\, R^2(E, \alpha_m)\ . \tag{2}$$

Here $A_{\text{ap},m}$ is the geometric (energy-independent) area of the mirror shell's aperture and $R(E, \alpha_m)$ is the (energy-dependent) reflectance at a grazing angle equal to the axial slope (cone angle) $\alpha_m$ of the primary mirror[h].

An unfortunate but necessary feature of grazing-incidence reflection (§2.2) is that the mirror surface area $A_{\text{surf}}$ is very much larger than the aperture area: $A_{\text{surf}} \cong 2 A_{\text{ap}} / \sin\alpha$, where the factor of two occurs because there are 2 mirror surfaces per shell. Consequently, the area ratio $(A_{\text{surf}} / A_{\text{ap}})$ is roughly of order 100 for soft-x-ray telescopes and even larger for hard-x-ray telescopes.

### 2.4.3. Detection sensitivity

While finer angular resolution improves sensitivity by reducing detector noise (§2.4.1) due to (non-imaged) x-ray and charged-particle background, larger collecting area (§2.4.2) improves sensitivity by increasing detector signal due to the imaged x-ray source. Observations for which the (non-imaged) background dominates the noise are called "background-limited". Although observations with non-focusing x-ray telescopes are generally background-limited, only very long observations are background-limited for high-resolution x-ray telescopes. For typical observations with such telescopes, counting (Poisson) statistics dominate the noise and the observations are termed "photon-limited". Improving the sensitivity of photon-limited observations requires more collecting area (or longer exposures). For focusing telescopes with very large collecting areas but inadequate angular resolution, imaged x rays from adjacent cosmic sources spill over into the detection cell and dominate the noise within it. Such observations are said to be "confusion-limited", in that photons of one source are confused with those of another. Improving the sensitivity of confusion-limited observations requires finer angular resolution.

The ideal next-generation x-ray observatory (§4.2) would have an angular resolution as good as or better than that of the *Chandra X-ray Observatory*, with a collecting area at least two-orders-of-magnitude larger. There are substantial challenges—technological and programmatic—to realizing such an observatory (§5). However, addressing some of the key scientific questions in astronomy and cosmology requires it.

## 3. COSMIC X-RAY SOURCES

With the dramatic improvement in sensitivity afforded by high-resolution x-ray telescopes, it became evident that nearly every category of celestial object emits x rays. Detected x-ray sources include solar-system objects (planets, comets, and solar corona), stellar coronae, accretion disks and jets associated with compact stars (stellar-mass black holes, neutron stars, and white dwarfs) and with super-massive black holes in galactic nuclei, isolated neutron stars, exploding stars and their remnants, and hot gas in galaxy clusters. Relevant spectral-line mechanisms include charge exchange in partially ionized plasmas, fluorescence or radiatively excited atomic transitions, and thermally excited atomic transitions from hot plasmas; relevant continuum mechanisms include thermal bremsstrahlung (free-free emission) from hot ionized plasmas, blackbody emission from very hot stellar surfaces, synchrotron emission from relativistic electrons in magnetic fields, and Compton scattering from relativistic electrons immersed in a lower-energy radiation environment.

---

[h] The right-most expression assumes an optimized telescope design, for which the cone angle (axial slope) $\alpha_S$ of each secondary mirror is 3 times the cone angle $\alpha_P$ of its corresponding primary mirror. With such a design, which is invariably chosen, the mean grazing angle for an on-axis incident ray is the same for each of the two reflections and nearly equal to the mean axial slope (cone angle) of the primary mirror— $\vartheta_S = \vartheta_P = \alpha \equiv \alpha_P = \alpha_S / 3$.

This section gives a cursory survey of the x-ray universe, featuring x-ray images[i] obtained with the *Chandra X-ray Observatory*. In providing a sampling of cosmic x-ray sources, it illustrates the importance of high angular resolution.

Figure 3 is a mosaic of 30 x-ray-color images of the central region of our Galaxy. At the 26,000-light-year[j] (26-kly) distance to the Galactic Center, the mosaic spans a region approximately 900 ly by 400 ly. The Galactic Center is obviously a very crowded field, which requires *Chandra*'s arcsecond resolution to elucidate structure and to detect the numerous (≈1000) faint discrete sources in the presence of several bright sources and diffuse x-radiation from very hot (tens-of-million-degree) plasma ejected and shock heated by stellar explosions and winds. The bright source (Sagittarius A) near the center of the mosaic is associated with a super-massive ($\approx 3\times 10^6$ solar-mass) black hole at the nucleus of the Galaxy. The bright extended source at the far left is a supernova remnant. Most of the other bright sources are "binary x-ray sources", powered by accretion of gas onto a compact star from a binary-companion normal star.

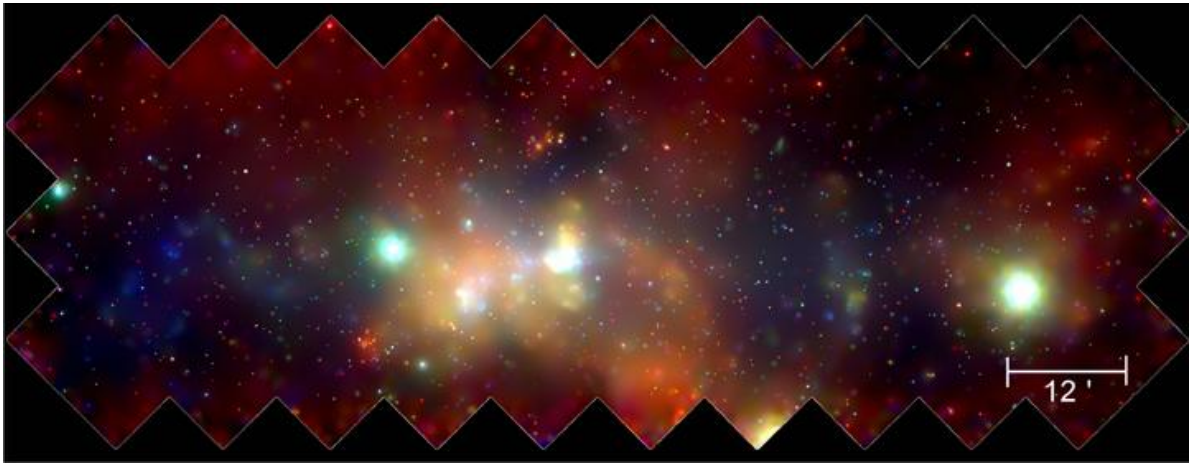

Figure 3: A mosaic image, spanning 2° × 0.8° in 30 pointings, of the (Milky Way) Galactic Center region in x-ray colors—1–3 keV (red), 3–5 keV (green), 5–8 keV (blue). [Credit: NASA/UMass/ Q. D. Wang et al.[51]]

Figure 4 displays images of two different types of supernova remnants (SNR), the aftermath of an exploding star (supernova). Cassiopeia A (left image, in x-ray colors) is a "shell-type" SNR, which is powered by the original explosion that occurred approximately 300 years ago. The supernova's blast-wave ejecta interacts with the circumstellar and interstellar media, producing forward and reverse shocks that heat the gas to (millions-of-degree) x-ray emitting temperatures. The thin, outer x-ray shell results from synchrotron emission from electrons in the interstellar medium that are shock accelerated to relativistic energies. *Chandra*'s arcsecond imaging not only reveals the intricate structure of the shocks and filaments, but also discovered the relic neutron star lying near the center of the SNR.

The Crab Nebula (right image, in composite colors) is a "filled-type" (*plerionic*) SNR, which is powered by the central pulsar (rapidly rotating, highly magnetic neutron star). In the resulting pulsar-wind nebula, relativistic electrons emit x rays profusely via the synchrotron mechanism. A remnant of a supernova explosion observed in 1054, the Crab Nebula is nearly 1000 years old. Over this time, the highest energy electrons radiate away their energy, which explains why the x ray nebula is significantly smaller than the nebular size observed in lower energy radiation. As already discussed (§2.4.1 and Figure 2), *Chandra*'s sub-arcsecond resolution manifests the complex structure of the pulsar-wind nebula.

---

[i] These and many more arcsecond-resolution x-ray images are available from the Chandra X-ray Center and may be accessed at http://chandra.harvard.edu/,

[j] A light year (ly), is of course the distance traveled by light in 1 year—$9.46\times 10^{15}$ m. A comparable distance unit commonly used by astronomers is the (parallax-second) parsec (pc), where 1 pc = 3.26 ly = $3.09\times 10^{16}$ m is the distance at which 1 astronomical unit (AU) subtends an angle of 1″ (4.848 μrad). The diameter of the earth's orbit around the sun defines the astronomical unit: 1 AU = $1.50\times 10^{11}$ m. The following guideline gives an indication of relevant astronomical scales: The separation between stars (in the solar neighborhood) is roughly a pc; the diameter of a galaxy, tens of kpc; the separation of galaxies, Mpc; and cosmologically significant distances, Gpc.

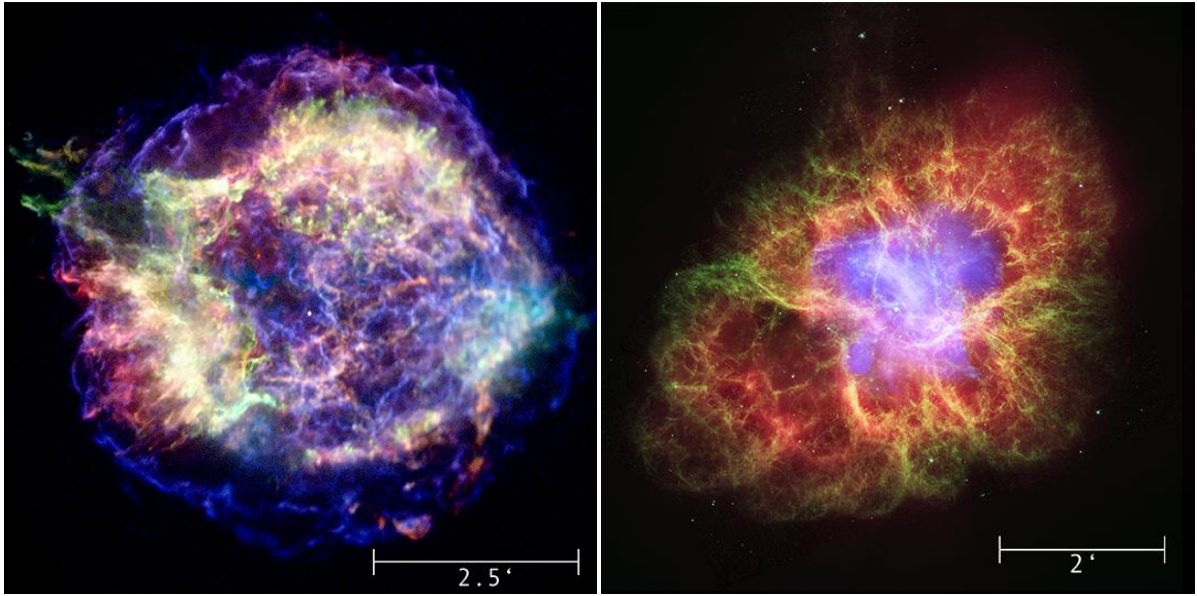

Figure 4: Supernova remnants (SNRs). The left image shows the shell-type SNR Cassiopeia A (Cas A) in x-ray colors—0.5–1.5 keV (red); 1.5–2.5 keV (green); 4–6 keV (blue). [Credit: NASA/CXC/MIT/UMass/ M. D. Stage et al.[52]] The right image displays the filled ("plerionic") SNR Crab Nebula in composite colors—x-ray (blue), visible (green), infrared (red). [Credit: (x-ray) NASA/CXC/ASU/ J. Hester et al; (visible) NASA/ESA/ASU/ J. Hester & A. Loll; (infrared) NASA/JPL-Caltech/UMinn/ R. Gehrz]

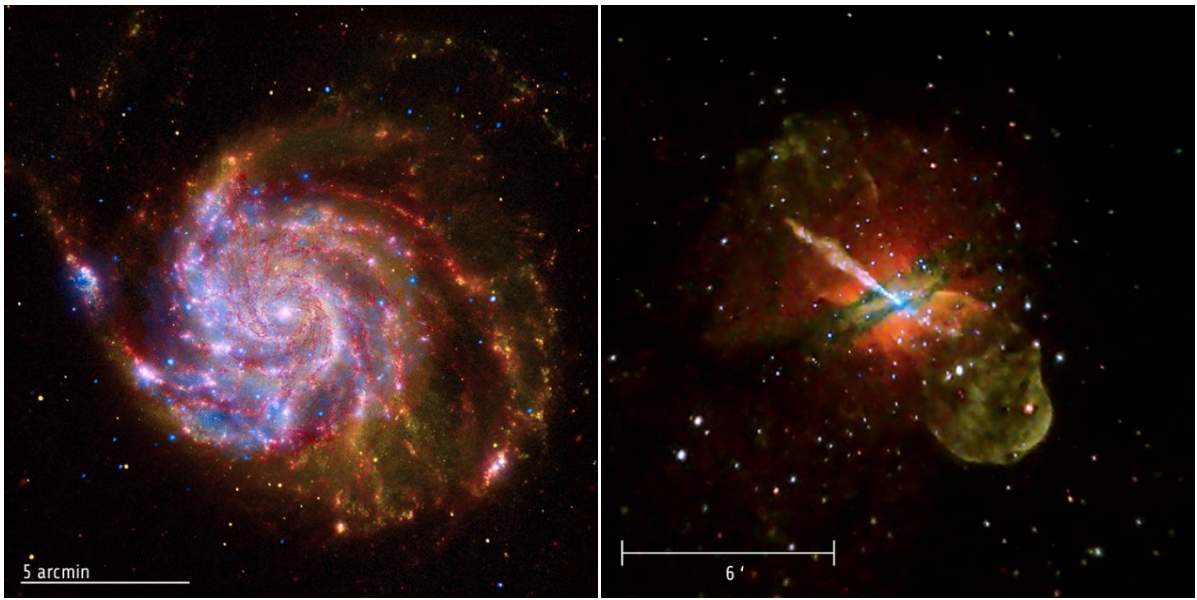

Figure 5: Galaxies. The left image shows the ("Pinwheel") spiral galaxy Messier 101 (M101) in composite colors—x-ray (blue), visible (yellow), infrared (red). [Credit: (x-ray) NASA/CXC/JHU/ K. Kuntz et al.; (visible) NASA/ESA/STScI/JHU/ K. Kuntz et al; (infrared) NASA/JPL-Caltech/STScI/ K. Gordon] The right image displays the active galaxy Centaurus A (Cen A) in x-ray colors—0.5–1.0 keV (red); 1.0–1.5 keV (green); 1.5–2.0 keV (blue). [Credit: NASA/CXC/CfA/ R. Kraft et al.]

Figure 5 exhibits images of the normal spiral Pinwheel Galaxy, Messier 101 (M101; left image, in composite colors) and of the active galaxy Centaurus A (right image, in x-ray colors). At a distance of about 22 Mly, M101 is similar to our Milky Way Galaxy (Figure 3), with its x-ray image exhibiting bright discrete sources (x-ray binaries and SNRs) and diffuse emission from hot gas (shock heated by stellar explosions and winds). Chandra's superb angular resolution enables the study of such crowded fields, which coarser-resolution telescopes would find quite confused (§2.4.3).

Centaurus A (right image, in x-ray colors) is a nearby (11-Mly distant) active galaxy, powered by accretion onto a super-massive ($\approx 10^8$ solar-mass) black hole in its nucleus. Prominent in the x-ray image, is the 13-kly-long primary jet and a shorter counter jet, which originate from the vicinity of the super-massive black hole. The image also shows hundreds of discrete sources (mostly x-ray binaries) and diffuse soft-x-ray emission, as are present in "normal" spiral galaxies—such as the Milky Way (Figure 3) and Pinwheel (Figure 5, left image) galaxies—with continuing star formation. The Cen-A dust lane, which is so prominent in visible light images, absorbs the low-energy x rays as well. Elucidating such complex structure requires *Chandra*'s fine angular resolution.

Figure 6 shows images of two clusters of galaxies—the largest gravitationally bound structures in the universe. Most of the baryonic (normal) mass in clusters of galaxies is in hot, diffuse x-ray-emitting gas, trapped in the gravitational potential of the cluster. Various lines of argument (based upon both x-ray and non-x-ray data) establish that most of the mass in clusters and in the universe as a whole is "dark"—i.e., non-baryonic. The convincing body of evidence—from microwave, visible, and x-ray observations over the past 20 years—for the existence and measurement of dark matter and dark energy, constitutes perhaps the greatest discovery in astronomy and cosmology since the early-twentieth-century formulation of General Relativity and observations showing that the universe is expanding. Indeed, the discovery of dark matter and dark energy may be even more important, in that it requires radically new physics.

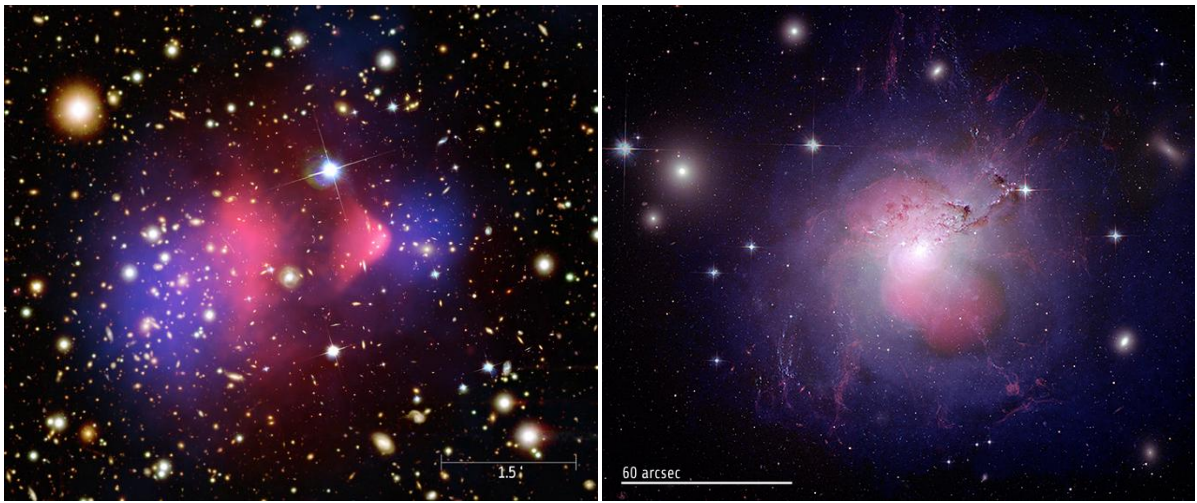

Figure 6: Clusters of galaxies. The left image displays the colliding ("Bullet") cluster 1E 0657-56 in composite colors—x-ray (pink); visible (orange/white); lensing (blue). [Credit: (x-ray) NASA/CXC/CfA/ M. Markevitch et al.; (visible) NASA/STScI/Magellan/UAriz/ D. Clowe et al.; (lensing) NASA/STScI/ESO/Magellan/UAriz/ D. Clowe et al.] The right image shows the Perseus Cluster in composite colors—x-ray (blue), visible (yellow/white), radio (red). [Credit: (x-ray) NASA/CXC/IoA/ A. Fabian et al.; (visible) NASA/ESA/STScI/IoA/ A. Fabian; (radio) NRAO/VLA/ G. Taylor]

The Bullet Cluster (left image, in composite colors) is actually two clusters of galaxies which have collided. The distribution of the thermal x-ray emission shows that gas in one cluster has interacted with that in the other: Hence, the two original gas clouds are combining to form a single cloud of hot plasma. On the other hand, the map of the matter distribution—as deduced from weak gravitational lensing[k]—shows two distinct mass concentrations after the collision: Consequently, most of the matter in the two original clusters did not interact during the collision. This is one of several lines of evidence for the existence and measurement of dark matter.

---

[k] In weak gravitational lensing, the mass of the cluster bends the light from background sources as the light ray passes near the cluster. This bending produces achromatic aberration (distortion) of the images of background sources, tending to elongate them in the direction tangential to direction of the mass concentration. Analysis of the shapes of these images then determines the distribution of matter transverse to the line of sight.

The Perseus Cluster (right image, in composite colors) is a rich cluster of galaxies, at a distance of about 250 Mly. The central dominant galaxy in the cluster is the active galaxy NGC 1275, whose super-massive black hole produces the strong, double-lobe radio source Perseus A. The composite-color image clearly shows that the radio synchrotron emission (from relativistic electrons) and the x-ray thermal emission (from hot plasma) originate in distinctly different regions. Indeed, the pressure within the radio lobes has pushed the thermal (x-ray-emitting) gas out of its way as it expanded into the cluster gas, producing holes in this cloud of thermal plasma.

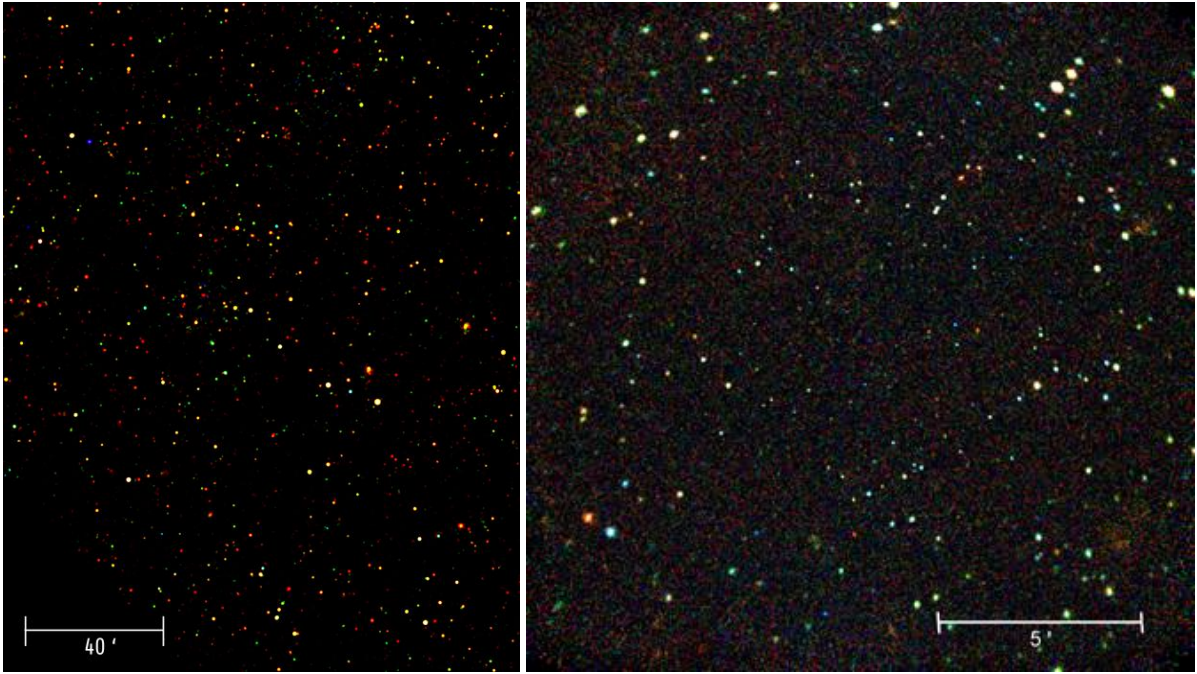

Figure 7: Surveys and cosmology. The left image shows the Boötes wide-field panorama, covering 9.3-square-degree in 126 pointings, in x-ray colors—0.5–1.3 keV (red); 1.3–2.5 keV (green); 2.5–7 keV (blue). [Credit: NASA/CXC/CfA/ R. Hickox et al.] The right image displays the (0.11-square-degree) Chandra Deep Field South in x-ray colors——0.3–1.0 keV (red); 1.0–2.0 keV (green); 2.0–7 keV (blue). [Credit: NASA/JHU/AUI/ R. Giacconi et al.]

Figure 7 displays the x-ray fields of two surveys conducted with the *Chandra X-ray Observatory*. The Boötes field (left image, in x-ray colors) is a wide-field (9.3-square-degree), medium-deep survey. The image combines 126 *Chandra* 5-ks exposures to cover so wide a field. Most of the detected x-ray sources in the field are active galactic nuclei (AGN), about half of which are obscured[53]— probably by a torus around the central super-massive black hole.

The Chandra Deep Field South (right image, in x-ray colors) is a narrow-field (0.07-square-degree), very-deep survey. This image combines observations of the field for a 1-Ms total exposure[54], which has subsequently been supplemented to a total 2-Ms exposure[55]. As with the less-deep x-ray surveys, most of the detected sources are (unobscured or obscured) AGN, but on-average more distant (higher redshift). The very deep survey also detects clusters of galaxies and relatively local less-luminous sources—galaxies and groups of galaxies. Achieving such extremely sensitive, deep surveys requires *Chandra*'s arcsecond resolution.

## 4. HIGH-RESOLUTION X-RAY OBSERVATORIES

The advent of high-resolution x-ray observatories revolutionized x-ray astronomy (§2.1). With the dramatically improved sensitivity afforded by focusing x-ray telescopes, x-ray astronomy expanded to include the study of nearly every category of celestial object. This, combined with the paradigm shift from principal-investigator experiments to general-observer facilities (pioneered by the *Einstein Observatory*), brought x-ray astronomy to the broader astronomy and astrophysics community.

This section briefly describes x-ray astronomy missions using high-resolution x-ray telescopes. To facilitate the focus on fine angular resolution, this discussion considers only x-ray telescopes with half-power diameter (HPD, §2.4.1) less than 15″. Section 4.1 summarizes the x-ray-telescope properties of past and current missions; Section 4.2, of future missions.

### 4.1. Past and current missions

Table 1 summarizes the properties of past and current high-resolution x-ray telescopes. The columns of the table give (1) the name of the mission and its mirror assembly, (2) the years of operation, (3) the telescope's focal length, (4) the diameter of the largest mirror shell, (5) the number of shells comprising the mirror assembly, (6) the mirror material, (7) the effective area at 1 keV, and (8) the telescope's angular resolution in terms of the approximate half-power diameter.

For completeness, the table includes (in the shaded rows) information on orbiting high-resolution x-ray telescopes for solar-physics research (*Yohkoh* SXT[56,57,58] and *Hinode* XRT[59,60]) and for space-weather monitoring (GOES SXI[61,62,63]). However, such solar x-ray telescopes are much smaller and less complex than those needed for x-ray astronomy. Indeed for solar x-ray telescopes a single, Wolter-1-like, monolithic, thick-walled mirror shell suffices. Consequently, the remainder of this section addresses only the (much larger) telescopes for x-ray astronomy.

Table 1: Past and current high-resolution x-ray telescopes with half-power diameter HPD < 15″.

| Mission<br>Telescope | Dates | Focal length [m] | Max diam. [m] | Shells [#] | Material | $A_{1\,keV}$ [m²] | HPD [″] |
|---|---|---|---|---|---|---|---|
| *Einstein Observatory* (HEAO-2)<br>High-Resolution Mirror Assembly | 1978–1981 | 3.4 | 0.56 | 4 | quartz | 0.02 | 10 |
| *Röntgen Satellit* (ROSAT)<br>X-Ray Telescope (XRT) | 1990–1999 | 2.4 | 0.83 | 4 | Zerodur™ | 0.04 | 5 |
| *Chandra X-ray Observatory*<br>High-Resolution Mirror Assembly | 1999– | 10.0 | 1.20 | 4 | Zerodur™ | 0.08 | 0.6 |
| XMM-*Newton*<br>(3 telescopes) | 1999– | 7.5 | 0.70 | 3 × 58 | electroform nickel | 3 × 0.14 | 14 |
| *Yohkoh* (Solar-A)<br>Soft X-ray Telescope (SXT) | 1991–2001 | 1.5 | 0.23 | 1 | Zerodur™ | 0.0001 | 5 |
| GOES<br>Solar X-ray Imager (SXI) | 2006– | 0.66 | 0.16 | 1 | Zerodur™ | 0.0002 | 4 |
| *Hinode* (Solar-B)<br>X-Ray Telescope (XRT) | 2006– | 2.7 | 0.34 | 1 | Zerodur™ | 0.0002 | 2 |

In many respects, the mirror assemblies for NASA's *Einstein Observatory*, Germany's ROSAT, and NASA's *Chandra X-ray Observatory* are quite similar: They each employed 4 co-axial Wolter-1 pairs[1] (paraboloid and hyperboloid) of thick-walled (≈ cm) mirrors, made of similar materials (fused quartz or the glassy ceramic Zerodur™). *Chandra* was the last and largest x-ray telescope using mechanically figured and super-polished, thick-walled "glass" grazing-incidence mirrors. As such, it profited from previous experience, particularly with the *Einstein Observatory*, and from a substantial development program for mirror technology. The result was spectacular—precision figured and super-polished mirrors that are precision aligned to achieve sub-arcsecond angular resolution.

---

[1] The original *Chandra* telescope design utilized 6 mirror pairs, which quantity was reduced to 4 shells in 1992. The main reason for deleting 2 shells was to reduce mass as part of a radical redesign and lightweighting of the spacecraft, in order to descope the mission from a serviceable near-earth orbit to a non-serviceable highly elliptical orbit. Interestingly, the 4 *Chandra* mirror shells are still numbered (largest to smallest) "1", "3", "4", and "6".

The design of ESA's XMM-*Newton* was complementary to that of *Chandra*. In order to ensure fine angular resolution, *Chandra* uses rather thick (≈ 2 cm) "glass" mirrors, which limits the collecting area consistent with mass constraints on the satellite. On the other hand, XMM-*Newton* uses rather thin (≈ 1 mm) electroformed-nickel replicated mirrors in order to achieve a large collecting area—albeit at the expense of angular resolution.

### 4.2. Future missions

The challenge in developing future x-ray observatories is to identify viable technologies for achieving both fine angular resolution and very-large collecting area in the same telescope. Figure 8 displays (to approximate scale) past, current, and future missions.

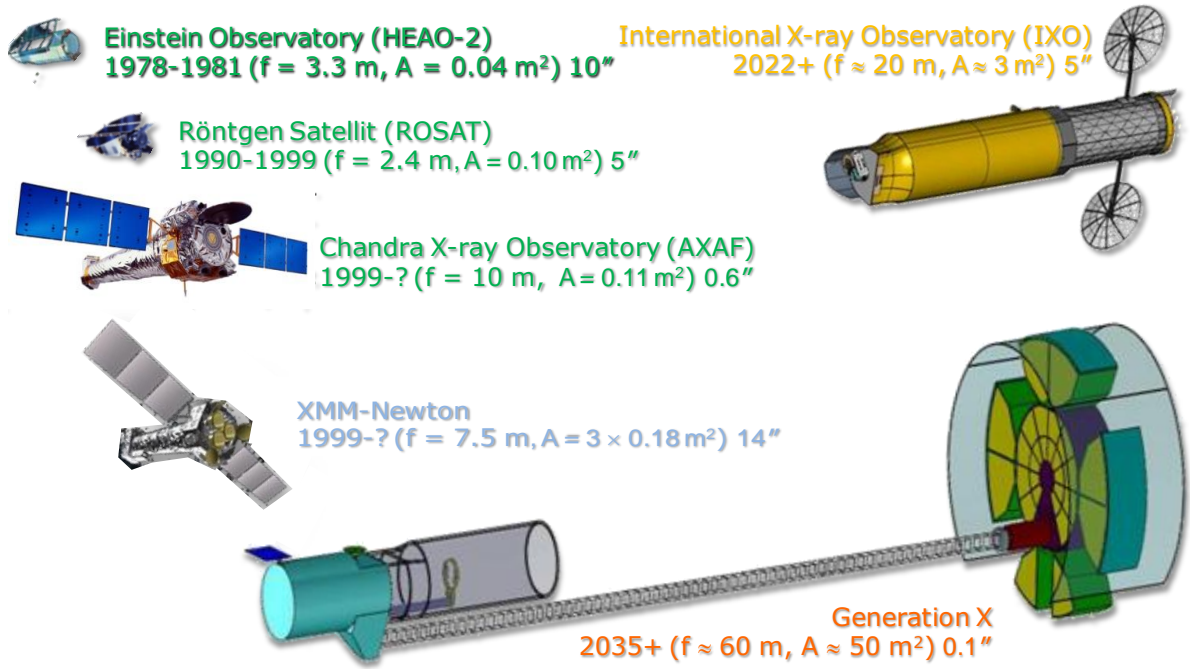

Figure 8: High-resolution x-ray observatories—past, present, and future[m]. Beside the image of each observatory are its dates of operation and the focal length $f$, aperture area $A$, and half-power diameter HPD of its x-ray telescope.

The *International X-ray Observatory* (IXO)—now under joint study by NASA, ESA, and JAXA—is to be the next-generation facility-class x-ray observatory, with a collecting area about an order of magnitude larger than that of XMM-*Newton* and an angular resolution 3 times finer. *Generation-X* (Gen-X)—an advanced concept studied for NASA—might be the next next-generation facility-class x-ray observatory, with an angular resolution finer than that of the Chandra X-ray Observatory and a collecting area a few-hundred times larger.

For any very-large-area telescope—such as IXO or Gen-X—the required collecting area is so large that it is impractical to construct a single mirror assembly comprised of full-cylinder mirror shells. One approach would be to build many smaller telescopes using full-cylinder shells: While this approach appears desirable from the standpoint of optics fabrication, it would require a large number of focal-plane detectors. The other approach—adopted for current IXO designs and Gen-X concepts—is to utilize segmented mirrors to synthesize cylindrical mirror shells.

---

[m] The *Astro2010* report "Decadal Survey of Astronomy and Astrophysics", which was released just 8 days after the conclusion of this SPIE Conference, recommends delaying the start of IXO mission development until the next decade. Implementation of this recommendation will likely delay the IXO launch until nearly 2030—i.e., about 30 years after the launches of the *Chandra X-ray Observatory* and of XMM-*Newton*.

IXO team members are developing two fundamentally different segmented-mirror technologies—slumped-glass optics (Figure 9) and silicon-pore optics (Figure 10). For slumped-glass optics, Goddard Space Flight Center (GSFC) leads the NASA-sponsored technology development[64]; INAF-Brera and Max-Plank-Institut fur extraterrestrische Physik (MPE) perform the ESA-funded technology development[65]. For silicon-pore optics, the European Space Research and Technology Centre (ESTEC) leads an industrial consortium in the ESA-sponsored technology development[66,67].

Figure 9 summarizes the steps required to fabricate an x-ray mirror assembly using the NASA approach for slumped-glass optics. Mirror production begins with the thermal slumping of a glass sheet over a precision-figured mandrel. Following removal of the slumped glass from the mandrel, the successive mirror-production steps are trimming of the slumped sheet, depositing an optical coating onto the substrate, and charactering the figure of the mirror. Assembly of a mirror module starts with use of a transfer mount to align the mirror for bonding into the mirror module that holds nested, co-aligned primary and secondary mirrors to form an azimuthal section of a nested, full-cylinder mirror assembly. Finally, the mirror modules are aligned to a common focus and secured into the mirror assembly, which is then integrated into the observatory.

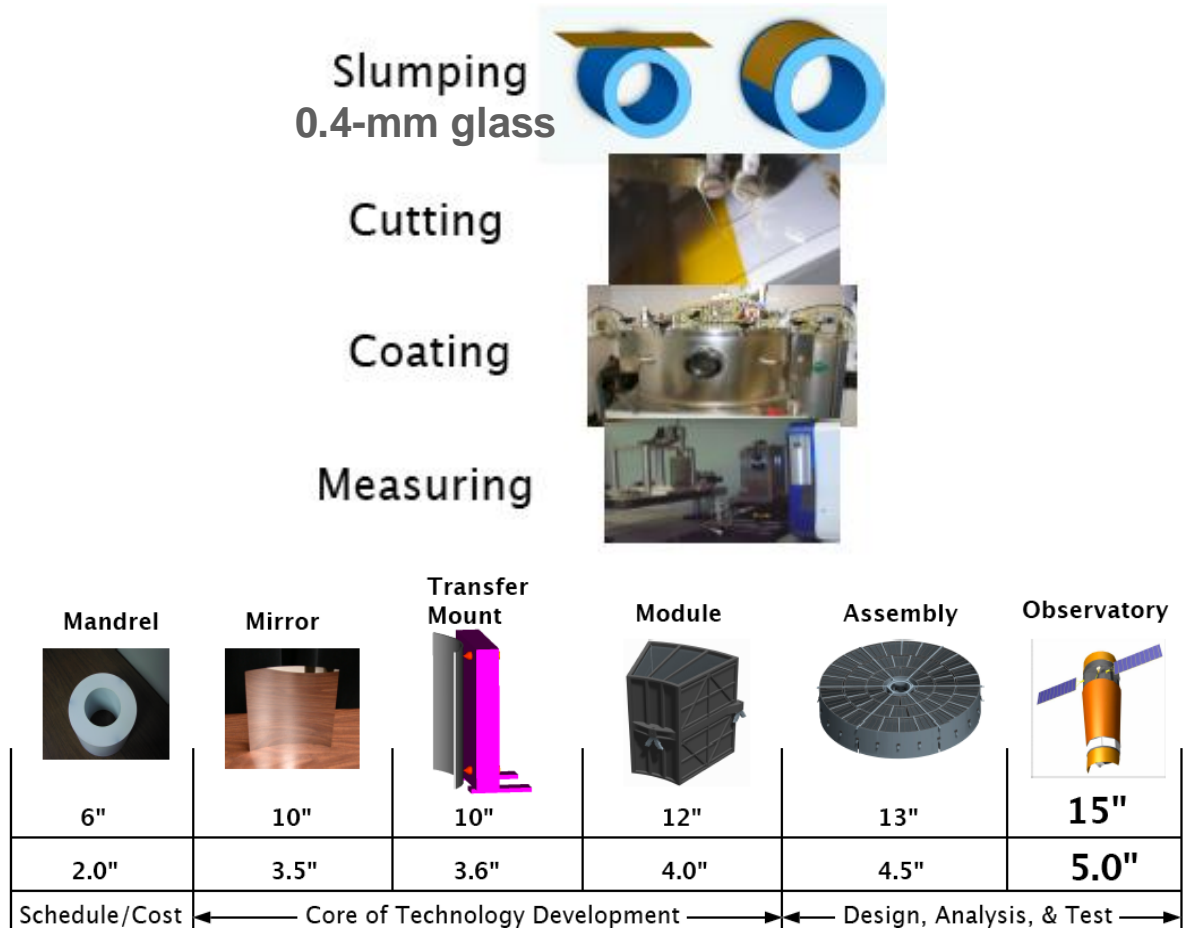

Figure 9: Slumped-glass optics. The top panel illustrates mirror-fabrication processes—thermally slumping glass, trimming edges, coating, and mirror metrology. The bottom panel follows the integration of mirrors into a mirror module, mirror assembly, and then observatory, with an allocated accumulation of imaging blur for a total observatory HPD of 15″ or 5″. [Credit: NASA/GSFC/ W. Zhang]

The table at the bottom of Figure 9 estimates the imaging error build-up through the steps of mirror fabrication, alignment, assembly, and on-orbit operation. The upper line gives cumulative allocations for achieving the original IXO resolution requirement of 15″ HPD; the lower line, for achieving the current requirement of 5″ HPD.

Silicon-pore optics (SPO) employ a novel approach for fabricating very-stiff, segmented mirror modules. The fabrication procedures robotically stack smooth, specially processed silicon-plate mirrors into a stiff mirror module termed an "x-ray optical unit" (XOU), which are then aligned to a common focus to comprise the mirror assembly. In Figure 10, the left panel sketches the largely-automated steps followed in fabricating an XOU, such as that appearing in the right panel of the figure.

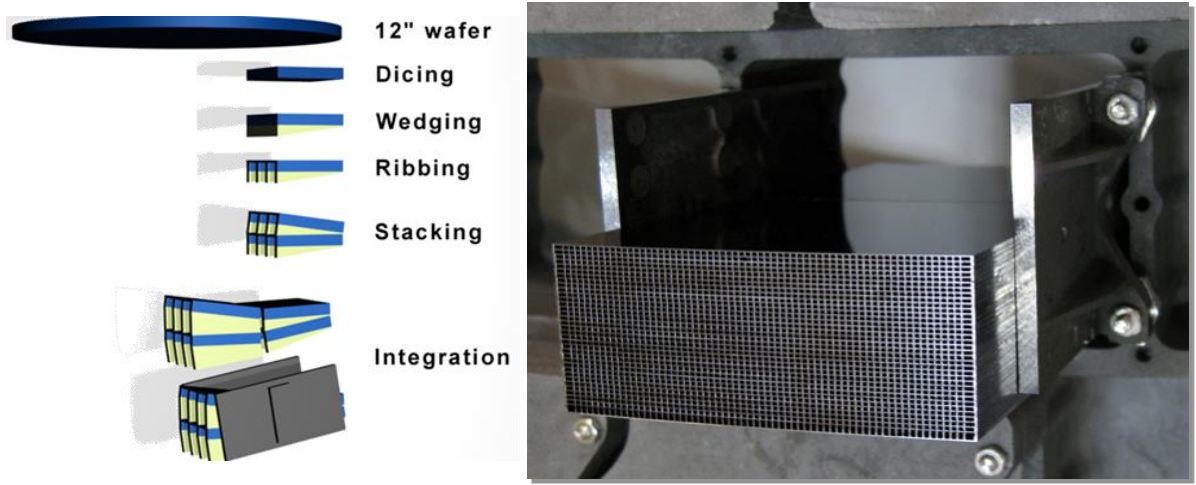

Figure 10: Silicon-pore optics. The left panel illustrates the process (from top to bottom) for fabricating an x-ray optical module (XOU)—selection of a silicon wafer, dicing into rectangular plates, wedging for grazing incidence, ribbing integral support structure, patterned coating (not shown), stacking wedged plates on a mandrel, and integration into an XOU. The right photograph displays an XOU with a 35-plate stack. [Credit: ESA/cosine/MPE/KT/SRON/Micronit/ M. Bavdaz]

Although the *International X-ray Observatory* is to have a collecting area about 30 times that of the *Chandra X-ray Observatory*, its angular resolution is expected to be about 10 times coarser. Being very ambitious, the *Generation-X* concept calls for a collecting area yet another order of magnitude larger than that of IXO and an angular resolution at least as good as *Chandra*'s, with a goal 5 times better. While neither of the mirror technologies being developed for IXO has yet satisfied the IXO angular-resolution requirement, each of them appears to be plausibly in reach of doing so within the next few years. In contrast, the technological and programmatic challenges (§5) to realizing a very-large-area, high-resolution x-ray telescope comparable to *Generation-X* are currently daunting.

## 5. CHALLENGES FOR A VERY-LARGE AREA, HIGH-RESOLUTION X-RAY TELESCOPE

Figure 11 compares key performance metrics (§2.4)—aperture area and half-power diameter—of past, current, and future high-resolution x-ray telescopes (§4). Improving the angular resolution of a telescope—i.e., moving to the left on this plot—is chiefly a technological challenge. Increasing the aperture area of a telescope—i.e., moving upward on this plot—is mainly a programmatic challenge.

For purposes of discussion, consider the flow-down of the performance requirements for the Generation-X mission concept. The key performance requirements are an angular resolution of HPD = HEW ≤ 0.1″ and an aperture area $A_{ap} \geq 50 \text{ m}^2$. The angular-resolution requirement flows down to the requirement that the mirror's RMS axial-slope deviations $\sigma_\alpha \leq 0.13$ μrad. While it is possible to figure thick mirrors to this accuracy, doing so for large mirror areas using current precision figuring techniques could be quite slow and expensive. For mean grazing angles typical of soft-x-ray telescopes, the mirror surface area is roughly 200 times the aperture area, thus the Gen-X aperture-area requirement flows down to a requirement that the total precision-figured and super-polished mirror area $A_{surf} \geq 100000$ m$^2$: This is a definite programmatic challenge! The technological challenge is to develop an approach for establishing and maintaining the prescribed figure and alignment of a lightweight mirror from ground to orbit and throughout the mission: One such approach may use adaptive x-ray optics to allow in-space adjustment of mirror figure and alignment.

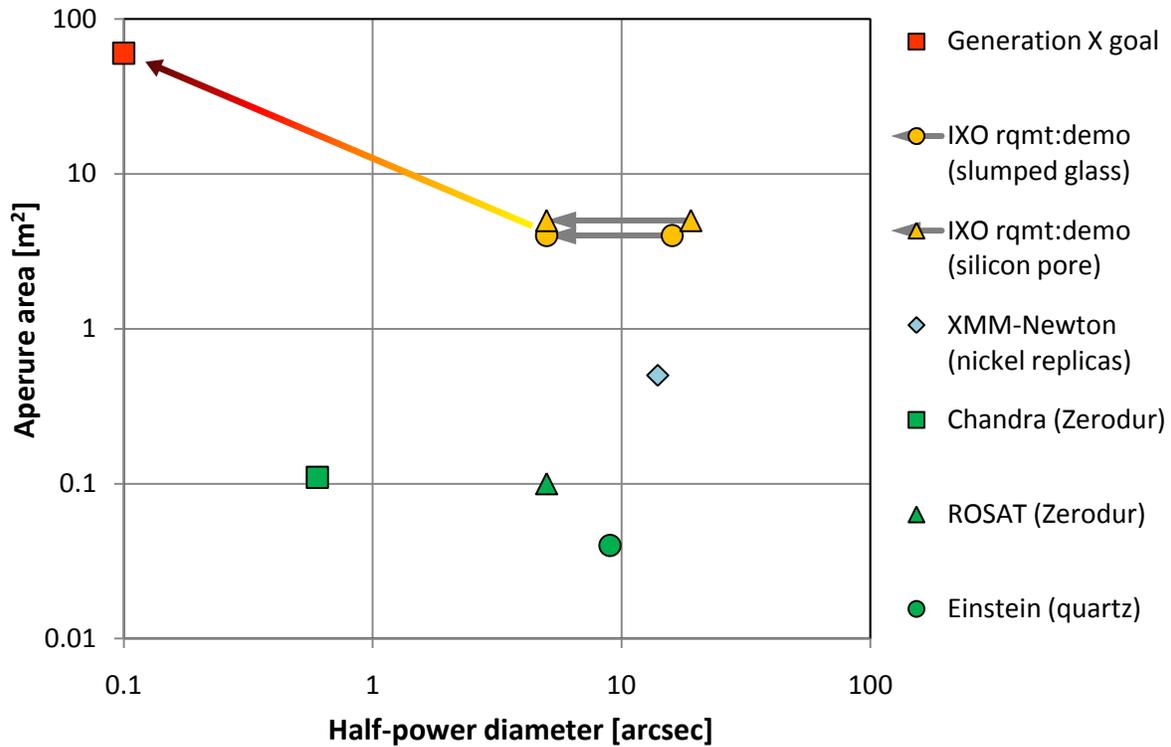

Figure 11: Comparison of performance metrics (aperture area and HPD resolution) of past, present, and future high-resolution x-ray telescopes. Achieving future scientific objectives will require more aperture area and eventually even better angular resolution.

Figure 12 summarizes requirements on the mirrors, flowed down from the imaging performance requirements. Combining these requirements with mass, monetary, and schedule constraints, limits the maximum allowable mirror areal density, maximum mirror areal cost, and minimum production rate. Satisfying these limits presents a daunting programmatic challenge, which will require significant technological innovation to overcome.

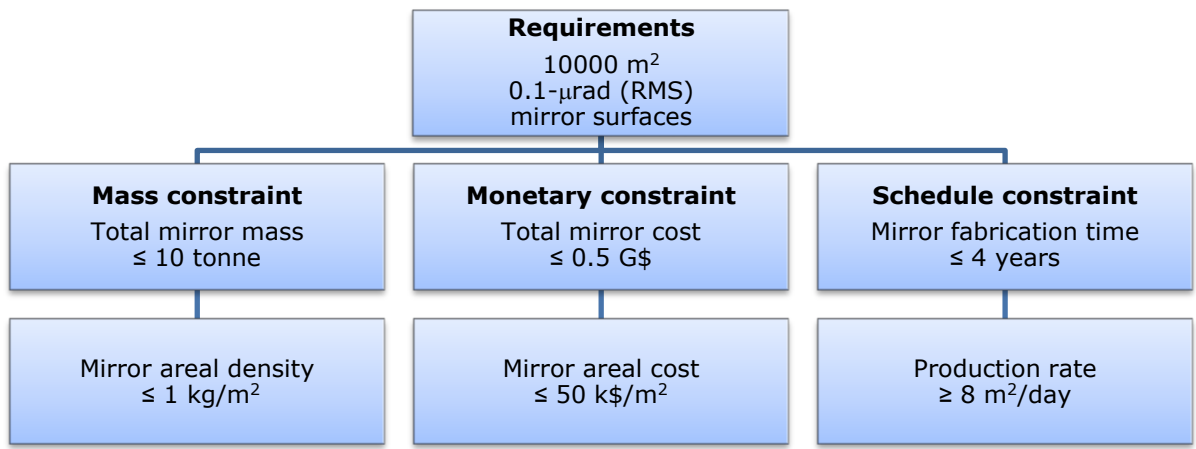

Figure 12: Technological and programmatic constraints for a next next-generation x-ray-astronomy facility-class mission. Fabricating and assembling very large areas of precision mirror surfaces will be technologically and programmatically challenging.

# ACKNOWLEDGEMENTS

The Gen-X work was conducted under NASA grant NNX08AT62G-R, "A Concept Study of the Technology Required for Generation-X – A Large Area High Angular Resolution Telescope".

# REFERENCES

1. Friedman, H., Lichtman, S. W., & Byram, E. T., "Photon counter measurements of solar x rays and extreme ultraviolet light", PhRv 83, 1025-1030 (1951).
2. Giacconi, R., Gursky, H., Paolini, F. R., & Rossi, B. B., "Evidence for x rays from sources outside the Solar system", PhRvL 9, 439-443 (1962).
3. Giacconi, R., "The high energy X-ray universe", PNAS 107, 7202-7207 (2010).
4. Weisskopf, M. C., Tananbaum, H. D., Van Speybroeck, L. P., & O'Dell, S. L., "Chandra X-ray Observatory (CXO): overview", SPIE 4012, 2-16 (2000).
5. Doel, P., Atkins, C., Thompson, S., Brooks, D., Yao, J., Feldman, C., Willingale, R., Button, T., Zhang, D., & James, A., "Large thin adaptive x-ray mirrors", SPIE 6705, 0M:1-8 (2007).
6. Feldman, C., Willingale, R., Atkins, C., Brooks, D., Button, T., Doel, P., James, A., Rodriguez Sanmartin, D., Smith, A., Theobald, C., Thompson, S., Wang, H., & Zhang, D., "First results from the testing of the thin shell adaptive optic prototype for high angular resolution x-ray telescopes", SPIE 7437, 1G:1-11 (2009).
7. Atkins, C., Doel, P., Brooks, D., Thompson, S., Feldman, C., Willingale, R., Button, T., Rodriguez Sanmartin, D., Zhang, D., James, A., Theobald, C., Smith, A. D., & Wang, H., "Advances in active x-ray telescope technologies", SPIE 7437, 1H:1-12 (2009).
8. Reid, P. B., Cameron, R. A., Cohen, L., Elvis, M., Gorenstein, P., Jerius, D., Petre, R., Podgorski, W. A., Schwartz, D. A., & Zhang, W. W., "Constellation-X to Generation-X: evolution of large collecting area moderate resolution grazing incidence x-ray telescopes to larger area high-resolution adjustable optic", SPIE 5488, 325-334 (2004).
9. Cameron, R. A., Bautz, M. W., Brissenden, R. J., Elvis, M. S., Fabbiano, G., Figueroa-Feliciano, E., Gorenstein, P., Petre, R., Reid, P. B., Schwartz, D. S., White, N. E., & Zhang, W. W., "Generation-X: mission and technology studies for an x-ray observatory vision mission", SPIE 5488, 572-580 (2004).
10. Reid, P. B., Murray, S. S., Trolier-McKinstry, S., Freeman, M., Juda, M., Podgorski, W., Ramsey, B., & Schwartz, D., "Development of adjustable grazing incidence optics for Generation-X", SPIE 7011, 0V:1-10 (2008).
11. Schwartz, D. A., Brissenden, R. J., Elvis, M., Fabbiano, G., Gaetz, T. J., Jerius, D., Juda, M., Reid, P. B., Wolk, S. J., O'Dell, S. L., Kolodziejczak, J. K., & Zhang, W. W., "On-orbit adjustment calculation for the Generation-X x-ray mirror figure", SPIE 7011, 0W:1-9 (2008).
12. Reid, P. B., Davis, W., O'Dell, S., Schwartz, D. A., Tolier-McKinstry, S., Wilke, R. H. T., & Zhang, W., "Generation-X mirror technology development plan and the development of adjustable x-ray optics", SPIE 7437, 1F:1-10 (2009).
13. Rodriguez-Sanmartin, D., Zhang, D., Button, T., Meggs, C., Atkins, C., Doel, P., Brooks, D., Feldman, C., Willingale, R., James, A., Willis, G., & Smith, A., "Development of net-shape piezoelectric actuators for large x-ray optics", SPIE 7803, 0M:1-8 (2010).
14. Feldman, C., Willingale, R., Atkins, C., Brooks, D., Button, T., Doel, P., James, A., Meggs, C., Rodriguez-Sanmartin, D., Smith, A., Theobald, C., & Willis, G., "The performance of thin shell adaptive optics for high angular resolution x-ray telescopes", SPIE 7803, 0N:1-10 (2010).
15. Reid, P. B., Davis, W., Schwartz, D. A., Trolier-McKinstry, S., & Wilke, R. H. T., "Technology challenges of active x-ray optics for astronomy", SPIE 7803, 0I:1-9 (2010).
16. Schwartz, D. A., Brissenden, R., Freeman, M., Gaetz, T., Gorenstein, P., Jerius, D., Juda, M., Reid, P., Wolk, S., Saha, T., Zhang, W., O'Dell, S., Trolier-McKinstry, S., & Wilke, R., "On-orbit adjustment concepts for the Generation-X Observatory", SPIE 7803, 0J:1-10 (2010).

17. Wilke, R. H. T., Trolier-McKinstry, S., Reid, P. B., & Schwartz, D. A., "PZT piezoelectric films on glass for Gen-X imaging", SPIE 7803, 0O:1-10 (2010).
18. Davis, W. N., Reid, P. B., & Schwartz, D. A., "Finite element analyses of thin film active grazing incidence x-ray optics", SPIE 7803, 0P:1-12 (2010).
19. Lillie, C., Pearson, D., Plinta, A., Metro, B., Lintz, E., Shropshire, D., & Danner, R., "Adaptive grazing incidence optics for the next generation of x-ray observatories", SPIE 7803, 0Q:1-7 (2010).
20. Proserpio, L., Civitani, M., Ghigo, M., & Pareschi, G., "Thermal shaping of thin glass substrates for segmented grazing incidence active optics", SPIE 7803, 0K:1-11 (2010).
21. Civitani, M., Ghigo, M., Citterio, O., Conconi, P., Spiga, D., Pareschi, G., & Proserpio, L., "3D characterization of thin glass x-ray mirrors via optical profilometry", SPIE 7803, 0L:1-13 (2010).
22. Tucker, W., & Giacconi, R., [The X-ray universe] Harvard University Press, Cambridge, MA USA (1985).
23. Bradt, H. V. D., Ohashi, T., & Pounds, K. A., "X-ray astronomy missions", ARA&A 30, 391-427 (1992).
24. Giacconi, R., "History of x-ray telescopes and astronomy", ExA 25, 143-156 (2009).
25. Tucker, W., "Major milestones in x-ray astronomy", http://chandra.harvard.edu/chronicle/0202/40years/index.html (2002).
26. Giacconi, R., & Rossi, B., "A 'telescope' for soft x-ray astronomy', JGR 65, 773-775 (1960).
27. Wolter, H., "Spiegelsysteme streifenden Einfalls als abbildende Optiken für Röntgenstrahlen" ("Grazing incidence mirror systems as imaging optics for x rays"), AnP 445, 94-114 (1952).
28. Giacconi, R, & Gursky, H., "An experimental program of extra-solar x-ray astronomy", American Science & Engineering (AS&E) proposal to NASA (1963).
29. Giacconi, R., Reidy, W. P., Zehnpfennig, T., Lindsay, J. C., & Muney, W. S., "Solar x-ray image obtained using grazing-incidence optics", ApJ 142, 1274-1278 (1965).
30. Giacconi, R., Harmon, N. F., Lacey, R. F., & Szilagyi, Z., "Aplanatic telescope for soft x rays", JOSA 55, 345-347 (1965).
31. Vaiana, G. S., Reidy, W. P., Zehnpfennig, T., Van Speybroeck, L., & Giacconi, R., "X-ray Structures of the Sun during the Importance 1N Flare of 8 June 1968", Sci 161, 564-567 (1968).
32. Underwood, J. H., & Muney, W. S., "A Glancing Incidence Solar Telescope for the Soft X-Ray Region", SoPh 1, 129-144 (1967).
33. Krieger, A., Paolini, F., Vaiana, G. S., & Webb, D., "Results from OSO-IV: the Long Term Behavior of X-Ray Emitting Regions", SoPh 22, 150-177 (1972).
34. Vaiana, G. S., Krieger, A. S., Petrasso, R., Silk, J. K., & Timothy, A. F., "The x-ray spectrographic telescope", SPIE 44, 185-205 (1974).
35. Walsh, E. J., Sokolowski, T. I., Miller, G. M., Cofield, K. L., Jr., Douglas, J. D., Lewter, B. J., Burke, H. O., & Davis, A. J., "Design characteristics of a SKYLAB soft x-ray telescope", SPIE 44, 175-184 (1974).
36. Giacconi, R., Branduardi, G., Briel, U., Epstein, A., Fabricant, D., Feigelson, E., Forman, W., Gorenstein, P., Grindlay, J., Gursky, H., Harnden, F. R., Henry, J. P., Jones, C., Kellogg, E., Koch, D., Murray, S., Schreier, E., Seward, F., Tananbaum, H., Topka, K., Van Speybroeck, L., Holt, S. S., Becker, R. H., Boldt, E. A., Serlemitsos, P. J., Clark, G., Canizares, C., Markert, T., Novick, R., Helfand, D., & Long, K., "The Einstein /HEAO 2/ X-ray Observatory", ApJ 230, 540-550 (1979).
37. Giacconi, R., & Tananbaum, H., "The Einstein Observatory: New perspectives in astronomy", Sci 209, 865-876 (1980).
38. Giacconi, R., Gorenstein, P., Murray, S. S., Schreier, E., Seward, F., Tananbaum, H., Tucker, W. H., & Van Speybroeck, L., "The Einstein Observatory and future X-ray telescopes", [Telescopes for the 1980s], Annual Reviews Inc., Palo Alto CA, 95-278 (1981).
39. Jerius, D., Donnelly, R. H., Tibbetts, M. S., Edgar, R. J., Gaetz, T. J., Schwartz, D. A., Van Speybroeck, L. P., & Zhao, P., "Orbital measurement and verification of the Chandra X-ray Observatory's PSF", SPIE 4012, 17-27 (2000).

40. Ramsey, B. D., Alexander, C. D., Apple, J. A., Benson, C. M., Dietz, K. L., Elsner, R. F., Engelhaupt, D. E., Ghosh, K., Kolodziejczak, J. J., O'Dell, S. L., Speegle, C. O., Swartz, D. A., & Weisskopf, M. C., "HERO: program status and first images from a balloon-borne focusing hard x-ray telescope", SPIE 4496, 140-145 (2002).

41. Ogasaka, Y., Tamura, K., Okajima, T., Tawara, Y., Yamashita, K., Furuzawa, A., Haga, K., Ichimaru, S., Takahashi, S., Fukuda, S., Kito, H., Goto, A., Kato, S., Satake, H., Nomoto, K., Hamada, N., Serlemitsos, P. J., Tueller, J., Soong, Y., Chan, K.-W., Owens, S. M., Berendse, F., Krimm, H., Baumgartner, W., Barthelmy, S. D., Kunieda, H., Misaki, K., Shibata, R., Mori, H., Itoh, K., & Namba, Y., "Development of supermirror hard x-ray telescope and the results of first observation flight of InFOCuS flight observation", SPIE 4851, 619-630 (2003).

42. Hailey, C. J., An, H., Blaedel, K. L., Brejnholt, N. F., Christensen, F. E., Craig, W. W., Decker, T. A., Doll, M., Gum, J., Koglin, J. E., Jensen, C. P., Hale, L., Mori, K., Pivovaroff, M. J., Sharpe, M., Stern, M., Tajiri, G., & Zhang, W. W., "The Nuclear Spectroscopic Telescope Array (NuSTAR): optics overview and current status", SPIE 7732, 0T:1-13 (2010).

43. Kunieda, H., Awaki, H., Furuzawa, A., Haba, Y., Iizuka, R., Ishibashi, K., Ishida, M., Itoh, M., Kosaka, T., Maeda, Y., Matsumoto, H., Miyazawa, T., Mori, H., Namba, Y., Ogasaka, Y., Ogi, K., Okajima, T., Suzuki, Y., Tamura, K., Tawara, Y., Uesugi, K., Yamashita, K., & Yamauchi, S., "Hard x-ray telescope to be onboard ASTRO-H", SPIE 7732, 14:1-12 (2010).

44. Joensen, K. D., Christensen, F. E., Schnopper, H. W., Gorenstein, P., Susini, J., Hoghoj, P., Hustache, R., Wood, J. L., & Parker, K., "Medium-sized grazing incidence high-energy x-ray telescopes employing continuously graded multilayers", SPIE 1736, 239-248 (1993).

45. Tawara, Y., Yamashita, K., Kunieda, H., Tamura, K., Furuzawa, A., Haga, K., Nakajo, N., Okajima, T., Takata, H., Serlemitsos, P. J., Tueller, J., Petre, R., Soong, Y., Chan, K.-W., Lodha, G. S., Namba, Y., & Yu, J., "Development of a multilayer supermirror for hard x-ray telescopes", SPIE 3444, 569-575 (1998).

46. Mao, P. H., Harrison, F. A., Windt, D. L., & Christensen, F. E., "Optimization of Graded Multilayer Designs for Astronomical X-ray Telescopes", ApOpt 38, 4766-4775 (1999).

47. Cotroneo, V., Bruni, R., Gorenstein, P., Pareschi, G., Romaine, S., Spiga, D., & Zhong, Z., "Hard x-ray reflectivity measurement of broad-band multilayer samples", SPIE 7437, 1Q:1-11 (2009).

48. Wolter, H., "Verallgemeinerte Schwarzschildsche Spiegelsysteme streifender Reflexion als Optiken für Röntgenstrahlen" ("Generalized Schwarzschild mirror systems of grazing reflection as optics for x rays"), AnP 445, 286-295 (1952).

49. O'Dell, S. L., Elsner, R. F., Kolodziejeczak, J. J., Weisskopf, M. C., Hughes, J. P., & Van Speybroeck, L. P., "X-ray evidence for particulate contamination on the AXAF VETA-1 mirrors", SPIE 1742, 171-182 (1993).

50. O'Dell, S. L., Elsner, R. F., & Oosterbroek, T., "Effects of contamination upon the performance of x-ray telescopes", SPIE 7732, 2V:1-16 (2010).

51. Wang, Q. D., Gotthelf, E. V., & Lang, C. C., "A faint discrete source origin for the highly ionized iron emission from the Galactic Centre region", Natur 415, 148-150 (2002).

52. Stage, M. D., Allen, G. E., Houck, J. C., & Davis, J. E., "Cosmic-ray diffusion near the Bohm limit in the Cassiopeia-A supernova remnant", NatPh 2, 614-619 (2006).

53. Hickox, R. C., Jones, C., Forman, W. R., Murray, S. S., Brodwin, M., Brown, M. J. I., Eisenhardt, P. R., Stern, D., Kochanek, C. S., Eisenstein, D., Cool, R. J., Jannuzi, B. T., Dey, A., Brand, K., Gorjian, V., & Caldwell, N., "A Large Population of Mid-Infrared-selected, Obscured Active Galaxies in the Boötes Field", ApJ 671, 1365-1387 (2007).

54. Giacconi, R., Zirm, A., Wang, J., Rosati, P., Nonino, M., Tozzi, P., Gilli, R., Mainieri, V., Hasinger, G., Kewley, L., Bergeron, J., Borgani, S., Gilmozzi, R., Grogin, N., Koekemoer, A., Schreier, E., Zheng, W., & Norman, C., "Chandra Deep Field South: The 1 Ms Catalog", ApJS 139, 369-410 (2002).

55. Luo, B., Bauer, F. E., Brandt, W. N., Alexander, D. M., Lehmer, B. D., Schneider, D. P., Brusa, M., Comastri, A., Fabian, A. C., Finoguenov, A., Gilli, R., Hasinger, G., Hornschemeier, A. E., Koekemoer, A., Mainieri, V., Paolillo, M., Rosati, P., Shemmer, O., Silverman, J. D., Smail, I., Steffen, A. T., & Vignali, C., "The Chandra Deep Field-South survey: 2 Ms source catalogs", ApJS 179, 19-36 (2008).

56. Lemen, J. R., Claflin, E. S., Brown, W. A., Bruner, M. E., & Catura, R. C., "Measurement of the point spread function and effective area of the Solar-A Soft X-ray Telescope mirror", SPIE 1160, 316-322 (1989).
57. Tsuneta, S., Acton, L., Bruner, M., Lemen, J., Brown, W., Caravalho, R., Catura, R., Freeland, S., Jurcevich, B., & Owens, J., "The soft X-ray telescope for the SOLAR-A mission", SoPh 136, 37-67 (1991).
58. Bruner, M. E., "The Soft X-ray Telescope for Solar-A - Design evolution and lessons learned", SPIE 1546, 222-264 (1992).
59. Deluca, E. E., "The X-Ray Telescope for Solar-B: Calibration and predicted performance", ASPC 369, 19-30 (2007).
60. Golub, L., Deluca, E., Austin, G., Bookbinder, J., Caldwell, D., Cheimets, P., Cirtain, J., Cosmo, M., Reid, P., Sette, A., Weber, M., Sakao, T., Kano, R., Shibasaki, K., Hara, H., Tsuneta, S., Kumagai, K., Tamura, T., Shimojo, M., McCracken, J., Carpenter, J., Haight, H., Siler, R., Wright, E., Tucker, J., Rutledge, H., Barbera, M., Peres, G., & Varisco, S., "The X-Ray Telescope (XRT) for the Hinode Mission", SoPh 243, 63-86 (2007).
61. Catura, R. C., Bruner, M. E., Catura, P. R., Jurcevich, B. K., Kam, C., Lemen, J. R., Meyer, S. B., Morrison, M. D., Magida, M. B., Reid, P. B., Harvey, J. E., & Thompson, P. L., "Performance of the engineering model x-ray mirror of the Solar X-ray Imager (SXI) for future GOES missions", SPIE 4138, 33-42 (2000).
62. Harvey, J. E., Atanassova, M., & Krywonos, A., "Systems engineering analysis of five 'as-manufactured' SXI telescopes", SPIE 5867, 0F:1-11 (2005).
63. Harvey, J. E., Krywonos, A., Atanassova, M., & Thompson, P. L., "The Solar X-ray Imager on GOES-13: design, analysis, and on-orbit performance", SPIE 6689, 0I:1-9 (2007).
64. Zhang, W. W., Atanassova, M., Augustin, P., Blake, P. N., Byron, G., Carnahan, T., Chan, K. W., Fleetwood, C., He, C., Hill, M. D., Hong, M., Kolos, L., Lehan, J. P., Mazzarella, J. R., McClelland, R., Olsen, L., Petre, R., Robinson, D., Russell, R., Saha, T. T., Sharpe, M., Gubarev, M. V., Jones, W. D., O'Dell, S. L., Davis, W., Caldwell, D. R., Freeman, M., Podgorski, W. A., & Reid, P. B., "Mirror technology development for the International X-ray Observatory mission", SPIE 7437, 0Q:1-12 (2009).
65. Ghigo, M., Basso, S., Bavdaz, M., Conconi, P., Citterio, O., Civitani, M., Friedrich, P., Gallieni, D., Guldimann, B., Martelli, F., Negri, R., Pagano, G., Pareschi, G., Parodi, G., Proserpio, L., Salmaso, B., Scaglione, F., Spiga, D., Tagliaferri, G., Terzi, L., Tintori, M., Vongehr, M., Wille, E., Winter, A., & Zambra, A., "Hot slumping glass technology for the grazing incidence optics of future missions with particular reference to IXO", SPIE 7732, 0C:1-12 (2010).
66. Wallace, K., Bavdaz, M., Gondoin, P., Collon, M. J., Günther, R., Ackermann, M., Beijersbergen, M. W., Olde Riekerink, M., Blom, M., Lansdorp, B., & de Vreede, L., "Silicon-pore optics development", SPIE 7437, 0T:1-9 (2009).
67. Collon, M. J., Günther, R., Ackermann, M., Partapsing, R., Vacanti, G., Beijersbergen, M. W., Bavdaz, M., Wille, E., Wallace, K., Olde Riekerink, M., Lansdorp, B., de Vrede, L., van Baren, C., Müller, P., Krumrey, M., & Freyberg, M., "Silicon-pore x-ray optics for IXO", SPIE 7732, 1F:1-9 (2010).